\documentstyle[12pt]{article}
\def\Ho#1{$H_o =#1\;\hbox{km s}^{-1}\hbox{Mpc}^{-1}$}

\title{An Alternative View of Flat Rotation Curves. II. The
Observations}

\author{D.S.L. Soares \\ Observat\'orio Astron\^omico da Piedade \\
Departamento de F\'{\i}sica, ICEx, UFMG --- C.P. 702 \\ 30161-970,
Belo Horizonte --- Brazil \\ {\small e-mail: dsoares@fisica.ufmg.br}}

\date{Jan. 1994}

\begin{document}

\maketitle

\begin{abstract}
The rotation curves of 20 spiral galaxies are examined in the light
of a toy model (Soares 1992) which has as the main feature the
assignment of a high M/L ratio ($= 30; H_o =50\;\hbox{km s}^{-1}
\hbox{Mpc}^{-1}$) to the visible matter. The observed rotation of all
galaxies can be accommodated without the assumption of a dark
halo.

Moreover, the suggestion is made that
the fact that almost all available rotational velocity measurements are
derived from emission lines emitted by galaxian gas (either neutral or
ionized) makes them inappropriate as tracers of the galaxy gravitational
potential. To account for that, the model introduces an effective
potential meant to describe the hydrodynamics inside a
gaseous disk.  The general morphology of the curves (i.e.,
the presence of a plateau in $V(r) \times r$) is interpreted
in this framework as a consequence of the hydrodynamical characteristics
of galaxian disks.

The Tully-Fisher relation is expressed in terms of model parameters
and used as an additional constraint in the process of fitting the
model to the observed rotation of the galaxies.
\clearpage
\centerline{\bf Resumen}

\medskip

Son investigadas las curvas de rotaci\'on de 20 galaxias espirales
usando el modelo propuesto por Soares (1992) que tiene como
ca\-racter\'\i stica principal a de atribuir una alta raz\'on M/L ($= 30;
H_o = 50\;\hbox{km s}^{-1} \hbox{Mpc}^{-1}$) para la materia visible.
La rotaci\'on observada de todas las galaxias pueden ser acomodadas sin
hacer la suposici\'on de un halo obscuro.

Sin embargo, se sugiere que el hecho de que casi todas las medidas disponibles
de velocidades de rotaci\'on son deducidas a partir de l\'\i neas
emitidas por el gas de las galaxias (neutro o ionizado) las hace
inadecuadas como indicadoras del potencial gravitacional. Para considerar
esto, el modelo introduce un potencial efetivo apropiado para describir
la hidrodin\'amica dentro del disco gaseoso. La morfologia general de
las curvas (i.e., la presencia de un plateau en $V(r) \times r$) es
interpretada, neste tratamiento, como una conseq\"uencia de las
caracter\'\i sticas hidrodin\'amicas de los discos galacticos.

La relaci\'on de Tully-Fisher se expresa en terminos de los pa\-ra\-metros
del modelo y es usada como una restricci\'on adicional en el proceso
de ajuste del modelo a la rotaci\'on observada de las galaxias.

\bigskip

{\bf Key Words:} spiral galaxies -- galaxian rotation -- dark matter

\end{abstract}

\section{Introduction}
The mass discrepancy in spiral galaxies is still an unsolved problem in
astrophysics.  Topical or review papers have been copiously written in
the last decade (e.g., Faber and Gallagher 1979; Rubin et al. 1982;
Whitmore et al. 1984; van Albada et al. 1985; van Albada and Sancisi 1986;
Trimble 1987; Sanders 1990; Broeils 1992; Ashman 1992; Persic and
Salucci 1993; etc) demonstrating renewed interest on the subject.

The orthodox explanation of  flat rotation curves of spiral galaxies
is that they are the result of combined effects of luminous matter, with M/L
typical of the solar neighborhood, and dark matter with much higher M/L.
Luminous matter accounts for the rotation inside the luminous body and a dark
extended halo is needed to fully account for the values above Keplerian
observed in the outer regions of spirals. Popular alternative views
include modifications of Newtonian dynamics (or gravity, e.g., Milgrom
1983, 1988. 1991), and modifications of the classical view of spiral
galaxies (e.g., Valentijn 1990; Gonz\'alez-Serrano and Valentijn
1991; Battaner et al. 1992). This paper follows another
(Soares 1992, hereafter Paper I) where we put forward an alternative
idea for the mass discrepancy query.

In Paper I we suggested that real
M/L for the luminous body of
spiral galaxies should be those obtained from binary galaxy
studies (e.g., Schweizer 1987; Soares 1989; etc), which are as large as
ten or fifteen times the local solar M/L. These values, in spite of
disagreeing with classical stellar population calculations (e.g., Larson
and Tinsley 1978), are by no means ruled out by standard stellar
population synthesis of spiral galaxies; it is almost
common sense nowadays that a population synthesis
calculation of spiral galaxies can yield virtually
{\it any} value of M/L, given the uncertainty on
stellar metallicities and ages.

Furthermore, we pointed out that
the observations of the rotation of spiral galaxies are inferred almost
always from Doppler shifts of emission lines generated either by
ionized or by neutral galaxian gas, and gas motion cannot be
used as a reliable tracer of the gravitational potential. Rather,
gas motion is ruled by an {\it effective } potential wherein
hydrodynamical effects are incorporated.

The last two paragraphs state the two basic work hypotheses of
the alternative model (hereafter AMOD) described in Paper I.

In section 2 a short summary of the toy model of Paper I is made.
The rotation curves of 20 spiral galaxies, observed
by Rubin et al. (1982) are investigated
in the framework of the AMOD potential (see equation (1), below).
In order to reduce the number of free parameters, we assume a fixed
M/L for all of the galaxies in our sample. The adopted value is
$M/L=30$ (for $H_o =50\;\hbox{km s}^{-1}\hbox{Mpc}^{-1}$,
here and throughout this paper). This
figure is consistent with binary galaxy studies and is further
justified by the application of the toy model to the well-known
spiral galaxy NGC3198 (van Albada et al. 1985; Begeman 1987, 1989; etc);
this is done in section 3. In section 4, an expression for the
Tully-Fisher relation (Tully and Fisher 1977), in terms of model
parameters, is derived, the fitting process is described and applied
to the sample. The resulting Tully-Fisher diagram is plotted
as well. A brief discussion of the results and our final conclusions are
presented in section 5.

\section{AMOD}
The overall kinematics of gas in spiral galaxy disks is derived
from the toy potential
$$U(r) = { {GM} \over {r} }
\left(1 + \beta e^{-r/r_{o}}\right)~~, \eqno(1)$$
where $\beta$ and $r_o$ are intrinsic galaxian scale parameters.

A Seeliger-Neumann type potential (North 1965; Assis 1992) was added to the
Newtonian gravitational potential in order to describe the hydrodynamics
inside a spiral gaseous disk; the parameters $\beta$ and $r_o$
determine the range of applicability of the new potential component
in each galaxy. In Paper I, this additional component
is associated with a kind of {\it buoyancy} potential. It is shown
there also that equation (1) can be derived from
a simple phenomenological description of the dynamics of gas
bubbles inside smooth gaseous disks.

The circular velocity of test particles is easily derived from
the AMOD potential:
$$v_{circ}(r) = {\left \{
{ {GM} \over {r} } \Bigl[ 1 + { \beta }
\Bigl(1 + {r \over r_{o}}\Bigr)e^{-r/r_{o}}\Bigr] \right \}
}^{1/2}~~. \eqno(2)$$
The Keplerian rotation curve derived from the light profile
of a spiral galaxy plus the usual values of M/L is below
observed gas rotation in the outer regions. The conventional
dark matter approach is to adopt an extended halo in such a
way that the model velocities are pushed up to the observed levels.
AMOD, on the other hand, assumes a higher M/L, resulting in
that the Keplerian curve will be located above the observed
rotation curve. With properly chosen $\beta$ and $r_o$ it was
shown in Paper I that equation (2) pulls down such an overestimated
rotation profile to the observed levels. This is illustrated
in Figure 1, where we plot the rotation curve of a conventional $M/L=3$
spiral galaxy. The Keplerian circular velocities are below
the observations and are reconciled with them with the aid of a dark halo.
Also shown is the Keplerian rotation curve of a $M/L=30$ spiral galaxy
that does not fit the observations as well. In this latter case the
AMOD potential is applied and the observations are now well fitted.
The exponential decay component present in the AMOD potential
represent an additional repulsive force of hydrodynamical
origin, which acts on the gas test particles that are
tracing the galaxy rotation.

\section{AMOD fit to NGC3198}
As we have pointed out in Paper I, AMOD has the same mathematical
formulation as the model suggested by Sanders (1984, 1986) to
account for mass discrepancies in spiral galaxies, i.e., both are
described by the same potential given by equation (1) with the sole
difference that, in Sanders' model, the gravitational constant $G$ is
replaced by a new $G_\infty$ (see below). This model,
named FLAG ({\it finite length-scale anti-gravity}), was later
abandoned by Sanders (e.g., 1990) mainly on the grounds of its inability
to explain one of the most important observed features of
spiral galaxies, namely the Tully-Fisher relation.

In FLAG, due to
the new potential component (the very same Seeliger-Neumann potential
shown above) the ``effective'' gravitational constant is now $G_\infty =
G/(1+\beta )$, where $\beta$, in the context of FLAG, is the coupling
constant of the additional component of gravity
\footnote {It is important
to stress that in AMOD such a component is not an
additional component of gravity, rather, it is a part of the
``effective'' galaxian potential, and is responsible for describing the
hydrodynamical behavior of gas test particles inside a gaseous disk.}
. $G$ is the usual Newtonian gravitational constant. Sanders
has used the galaxy NGC3198 to find out what the values of
FLAG's {\it universal} constants, $r_o$ and $\beta$, are. He
found, for a luminous disk with $M/L=2.4$,
$r_o = 36$ kpc and $\beta = -0.92$ (with \Ho{50}). These
values can be used to calculate the AMOD $M/L$ of NGC3198, i.e.,
$2.4/(1+\beta) = 30$.  They are also used to fit
the observed rotation curve of NGC3198 with equation (2).
Again, as in Paper I, we represent $M$ by a Plummer sphere (an $n=
5$ polytrope). Its cumulative mass distribution is given by
$$M(r) = {{M_o r^3} \over {(r^2+\epsilon^2)^{3/2}}}~~, \eqno(3)$$
where $M_o$ is the total galaxy mass and $\epsilon$ is the
Plummer sphere core radius. In order to account for the fact
that spirals have constant central surface brightness of
$\approx 140~ \hbox{L}_\odot \hbox{pc}^{-2}$
(Freeman 1970; Schweizer 1976), the core radius is scaled
to the total galaxy mass as (e.g., Sanders 1988):
$$\epsilon = 1.6 ~~M_o^{1/2}~~. \eqno(4)$$
$M_o$ is given in units of $10^{11}~ \hbox{M}_\odot$ and $\epsilon$
in kpc. The total luminosity (B band) of NGC3198 is $2.0 \times 10^{10}~
\hbox{L}_{B\odot}$ (Begeman 1987), which implies $M_o= 6.0 \times
10^{11}~ \hbox{M}_\odot$ and $\epsilon = 3.9 $ kpc. Equation (4)
is also used below (section 4) to model all the sample galaxies.

Figure 2 shows the fit to the observed rotation
curve of NGC3198 obtained with the above-mentioned parameters.
The only difference between the AMOD fit and FLAG's one is
the use by Sanders of an exponential disk to model the galaxy.
As one sees, with the spherical symmetric Plummer model the
fit does not change too much. The general behavior of the
rotation curve is reproduced in spite of losing fine features.

{}From the above investigation on NGC3198, and from results of binary
galaxy studies (Schweizer 1987; Soares 1989; etc), we shall adopt
the value of $M/L=30$ to proceed with the AMOD analysis. Of course,
the $M/L$ of a given galaxy is an intrinsic parameter and might
vary from galaxy to galaxy. The point here is that fixing $M/L$,
we gain in reducing the free parameter space while not being
too far from what ought to be individual values of $M/L$ for each
galaxy. It is worthwhile noting that, in the context of AMOD, such
a high $M/L$ is assigned to the {\it visible} body of the galaxy
instead of being justified by the artificial assumption of the
existence of an extended dark halo surrounding the visible
galaxy.

The sample of 20 Sb galaxies of Rubin et al. (1982) will
be considered in the next section taking into account the AMOD
potential and the Tully-Fisher relation.

\section{AMOD and the Tully-Fisher relation}
The main question in this section is whether AMOD is
consistent or not with the Tully-Fisher relation. To investigate
that, a set of observed rotational profiles is modeled with
AMOD having the Tully-Fisher relation as an additional fitting
constraint.

To begin with, one needs a convenient expression for the
Tully-Fisher relation, i.e., a relationship which involves the relevant
AMOD parameters. Let us consider the following form of the
Tully-Fisher relation (Aaronson, Huchra and Mould 1979):
$$L~=~k_{TF}~V_p^4~~, \eqno(5)$$
which, in spite of being oversimplified, is very useful to
the sort of investigation underway.

For the evaluation of $k_{TF}$ we have considered the spiral galaxies
observed in the 21 cm line of neutral hydrogen by Begeman (1987).
We took, for Begeman's eight galaxies, the luminosity in the B band
($L$) and the plateau velocity of the rotation
curve ($V_p$). From a plot of $L \times V_p^4$, and in
a system of units where luminosity is given in
$10^{11}~ \hbox{L}_\odot$, time in $10^8$ s,
velocity in units of 97.8 km/s and $G=4.50$,
we derive $k_{TF}=2.87~ \times 10^{-2}$. A range
of 30\% error is allowed in equation (5) (see below) to account
for the scattering in the diagram $L \times V_p^4$ which
was used for the determination of $k_{TF}$.

The AMOD parameters are introduced in equation (5) through the
expression for the circular velocity given by equation (2). Let
$V_p=v_{circ}(r=2\times R_{25})$, where $R_{25}$ is the
de Vaucouleurs' radius of the galaxy (i.e., the radius of
the galaxy to the 25th B magnitude per square arcsecond isophote).
This is an arbitrary value for the model plateau velocity but the
particular choice of $V_p$ (at some $r \geq R_{25}$) does not change
the analysis, as we have verified. The  galaxy mass is converted into
integrated blue luminosity via the $M/L$ ratio.  We can then re-write
equation (5) as
$$A={1 \over {Gk_{TF} V_p^2} }~~, \eqno(6)$$
where $A$ is the AMOD {\it index}, and is given by:
$$A={ {M/L \over 2R_{25}} } \Bigl[1+\beta\Bigl(1+{2R_{25}\over r_o}
\Bigr) e^{-2R_{25}/r_o}\Bigr]   ~~.   \eqno(7)$$
The Tully-Fisher relation is now expressed through a
relationship between the so-called AMOD index $A$
and $V_p$ (equation (6)). $A(V_p)$ is the AMOD function
one would expect if the Tully-Fisher relation can
be expressed as equation (5). Equation (7) can be used then to
derive an {\it observed} AMOD index ($A_{fit}$),
taking into account intrinsic galaxian parameters ($\beta,
r_o, R_{25}$ and $ M/L$). The AMOD parameters, $\beta$ and $r_o$,
are obtained from the fits to the rotation curves.

A fit of the AMOD circular velocity to Rubin et al. (1982)
observations of the rotation of 20 Sb spiral galaxies was done.
The fitting procedure took into account the goodness of fitting
to the Tully-Fisher relation (equation (6)) too. A quantitative measure
of the fitting of $A_{fit}$ (defined through equation (7)) to
equation (6) can be given by $\delta A$, defined by
$$ \delta A= { A(V_p)-A_{fit} \over 0.30A(V_p)  }~~.
\eqno(8)$$
This quantity gives the amount of deviation of the fitted
AMOD index from the predicted $A(V_p)$ curve in terms of
a 30\% error bar. The AMOD solutions ($V_{AMOD}(r)$) for
the rotation curves measured by Rubin et al. ($V_{obs}(r)$)
are those that simultaneously minimize $\mid\!\!\delta
A\!\!\mid$ and $\Delta V = \mid\!\! V_{obs}(r)- V_{AMOD}(r)\!\!\mid$.

Figure 3 and Figure 4 show the result of the fitting process.
The observed rotation curves by Rubin et al. are shown
together with the AMOD fit in Figure 3. The Tully-Fisher relation
is shown in Figure 4. Table 1 presents the relevant parameters
of all studied galaxies. Column 1: galaxy name (NGC or UGC);
column 2: plateau velocity, as given by Rubin et al. and by
Begeman in the case of NGC3198; column 3: de Vaucouleurs' radius,
in kpc; column 4: characteristic AMOD radius, in kpc; column
5: AMOD coefficient $\beta$; column 6: AMOD index, defined
by equation (7); column 8: relative error in $A$, defined by
equation (8).

Both trends, in the rotational profiles and in the
Tully-Fisher relation, are reasonably well reproduced,
as can be seen in Figures 3 and 4. It must be pointed out that
the fact that the fitting procedure consists in minimizing
$\mid\!\!\delta A\!\!\mid$ and $\Delta V$ does not necessarily
imply that the distributions of points in the
$A \times V_p$ diagram should follow the general trend
given by equation (6); the points could be distributed in any
possible way in that plane but, amongst all of them, they do follow
the behavior predicted from the observed Tully-Fisher
relation.

\section {Discussion and Conclusion}
Presently, it is a very difficult task to deny the existence of
extended dark material halos surrounding spiral galaxies, yet
accepting them remains as an uncomfortable position for many
astronomers. The main reason for that is of course the fact that
there is no {\it direct} observational evidence of such objects.
They must be very dark, indeed.

Here, we have investigated the possibility of avoiding the dark
halo hypothesis by means of a toy model which, in turn,
suffers from a serious setback: the whole idea is based upon a
drastic assumption in view of current standards, namely, it requires a
spiral galaxy with an exceedingly high $M/L$ (about $10-15$ times as large
as accepted values). On the other hand, and justifying
such a speculation, there is no firm {\it direct} observational evidence
supporting current M/L values of spiral galaxies.

A common feature of almost all model rotation curves, shown
in Figure 3, is the badness of fitting in the small radius range. The AMOD
potential is not able to give a detailed description of the rotational
properties of spiral galaxies in the inner regions. The reason for
that might be the oversimplified modeling of the hydrodynamical
behavior of the gas component. Nevertheless, it gives a
reasonable account of the global rotation of the gaseous medium
as can be seen in the fitting of all of the sample rotation curves.
It must be realized also that the Plummer sphere cannot
represent details of the luminosity (mass) profiles of real
galaxies, which has certainly contributed for
some of the features shown in Figure 3.

The main point we want to make here is that disk rotation curves
(either from gas emission lines or absorption from young stars)
are not reliable to trace out the gravitational potential of a
spiral galaxy as a whole. The toy model, discussed here and in Paper
I, suggests that gas rotation may be indicating internal properties
of the gaseous disk component rather than giving global information
about the galaxy mass distribution.

AMOD can be tested observationally through a comparison between
stellar and gas rotation curves in the inner regions of spiral
galaxies. There, the effects caused by the gaseous character
of the material are stronger than in the outer regions. That
could be done from absorption line measurements, in the case
of stellar rotation curves, and from emission line measurements
from HII regions. Of course, the gravitational potential
is the same for both gas and stars but the question is
whether the {\it effective} galactic potential for gas is the same
as for the stellar component. The physical processes acting on
the gas component are certainly different from those on the stellar
component. A special care must be taken when choosing the lines
for the absorption measurements: one should look preferentially
to old disk stars. Young OB stars preserve the kinematical
characteristics of the gas medium where they were formed and are
likely to give the same results for the rotation velocities
as compared to the gas measurements.

\vskip 1.5cm

{\noindent \it Acknowledgment --- } Partial support from CNPq (Conselho
Nacional de Desenvolvimento Cient\'\i fico e Tecnol\'ogico, Brazil
 --- Process No. 300193/90-4) is gratefully acknowledged. I would like to
thank Dr. F.O. V\'eas Letelier for the help in translating the abstract to
Spanish.

\vfill

\eject

{\noindent \bf References \\}
{
\noindent
Aaronson, M., Huchra, J., Mould, J.:
1979, {\it Astrophys. J.} {\bf 229}, 1 \\
Albada, T.S. van, Bahcall, J.N., Begeman, K., Sancisi, R.:
1985, {\it Astrophys. J.} {\bf 295}, 305 \\
Albada, T.S. van, Sancisi, R.: 1986, {\it Phil. Trans. Royal Soc. Lond.}
{\bf A 320}, 447 \\
Ashman, K.M.: 1992, {\it Publ. Astron. Soc. Pacific} {\bf 104}, 1109 \\
Assis, A.K.T.: 1992, {\it Apeiron} {\bf 13}, 3 \\
Battaner, E., Garrido, J.L., Membrado, M., Florido, E.: 1992, {\it
Nature} {\bf 360}, 624 \\
Begeman, K.G.: 1987, {\it Ph.D. Thesis}, University of Groningen \\
Begeman, K.G.: 1989, {\it Astron. Astrophys. } {\bf 223}, 47 \\
Broeils, A.H.: 1992, {\it Ph.D. Thesis}, University of Groningen \\
Faber, S.M., Gallagher, J.S.: 1979, {\it Ann. Rev. Astron. Astrophys. }
{\bf 17}, 135 \\
Freeman, K.C.: 1970, {\it Astrophys. J.} {\bf 160}, 811 \\
Gonz\'alez-Serrano, J.I., Valentijn, E.A.: 1991, {\it Astron.
Astrophys.} {\bf 242}, 334 \\
Larson, R.B., Tinsley, B.M.: 1978, {\it Astrophys. J.} {\bf 219}, 46 \\
Milgrom, M.: 1983, {\it Astrophys. J. } {\bf 270}, 365 \\
Milgrom, M.: 1988, {\it Astrophys. J. } {\bf 333}, 689 \\
Milgrom, M.: 1991, {\it Astrophys. J. } {\bf 367}, 490 \\
North, J.D.: 1965, {\it The Measure of the Universe -- A
History of Modern Cosmology}, Clarendon Press, Oxford \\
Persic, M., Salucci, P.: 1993, {\it Mon. Not. R. astr. Soc.} {\bf
261}, L21 \\
Rubin, V.C., Ford, W.K., Thonnard, N., Burstein, D.: 1982, {\it
Astrophys. J.} {\bf 261}, 439 \\
Sanders, R.H.: 1984, {\it Astron. Astrophys. } {\bf 136}, L21 \\
Sanders, R.H.: 1986, {\it Astron. Astrophys. } {\bf 154}, 135 \\
Sanders, R.H.: 1988, {\it Mon. Not. R. astr. Soc.} {\bf 235},105 \\
Sanders, R.H.: 1990, {\it Astron. Astrophys. Rev.} {\bf 2}, 1 \\
Schweizer, F.: 1976, {\it Astrophys. J. Suppl. Ser.} {\bf 31}, 313 \\
Schweizer, L.S.: 1987, {\it Astrophys. J. Suppl. Ser.} {\bf 64}, 427 \\
Soares, D.S.L.: 1989, {\it Ph.D. Thesis}, University of Groningen \\
Soares, D.S.L.: 1992, {\it Rev. Mexicana Astron. Astrof.} {\bf 24},
3 (Paper I) \\
Trimble, V.: 1987, {\it Ann. Rev. Astron. Astrophys.} {\bf 25}, 425 \\
Tully, R.B., Fisher, J.R.: 1977, {\it Astron. Astrophys. } {\bf 54}, 661 \\
Valentijn, E.A.: 1990, {\it Nature} {\bf 346}, 153 \\
Whitmore, B.C., Rubin, V.C., Ford, W.K.: 1984, {\it Astrophys. J.} {\bf 287},
66 \\
}

%\vfill

%\eject

\parbox{9.0cm}{\centerline{Table 1} }

\smallskip

{
\parbox{9.0cm}{Parameters and AMOD indices for
sample galaxies}
}

\medskip

\begin{tabular}{llllllc} \hline \hline
Name  & $V_p$ & $R_{25}$ & $r_o$ & $-\beta$& A & $\delta$A   \\
%    &(km/s) & (kpc) & (kpc) &     & & \\
(1) & (2) & (3) & (4) & (5) & (6) & (7) \\ \hline
N3200  & 282. &46.0 &140.0&0.90 & 0.74 & $-$0.68  \\
N7606  & 246. &39.8 &150.0&0.91 & 0.68 & $-$1.48   \\
N3223  & 255. &34.8 &120.0&0.92 & 0.80 & $-$0.98  \\
N3145  & 275. &33.8 &100.0&0.92 & 0.96 & $-$0.07  \\
& & & & & & \\
U12810 & 235. &51.4 &110.0&0.95 & 0.81 & $-$1.31  \\
N7083  & 222. &39.0 &120.0&0.92 & 0.80 & $-$1.56   \\
N1085  & 310. &34.8 &260.0&0.84 & 0.80 &  +0.12  \\
N2815  & 280. &26.5 &180.0&0.82 & 1.2  &  +0.85  \\
& & & & & & \\
N2590  & 256. &39.6 &100.0&0.88 & 1.1  & $-$0.14  \\
N3054  & 239. &25.7 & 60.0&0.90 & 1.7  &  +1.03  \\
N1620  & 252. &29.8 & 55.0&0.92 & 1.8  &  +1.72  \\
N7217  & 254. &15.1 &160.0&0.78 & 2.3  &  +3.37  \\
& & & & & & \\
N7537  & 144. &18.3 & 90.0&0.94 & 0.98 & $-$2.42   \\
N1325  & 184. &18.8 & 36.0&0.93 & 2.6  &  +0.69  \\
N1353  & 226. &14.0 &140.0&0.84 & 1.9  &  +0.97  \\
N1515  & 153. &13.6 &100.0&0.86 & 1.8  & $-$1.40  \\
& & & & & & \\
N4448  & 190. &11.0 &120.0&0.84 & 2.4  &  +0.49  \\
N3067  & 161. & 9.6 & 80.0&0.90 & 1.9  & $-$1.11  \\
N2708  & 241. &14.7 &120.0&0.86 & 1.7  &  +0.99  \\
N4800  & 179. & 4.9 & 48.0&0.84 & 5.4  &  +4.41  \\
& & & & & & \\
N3198  & 149. &17.1 & 36.0&0.92 & 2.7  & $-$0.65  \\ \hline
\end{tabular}

\medskip

{
\parbox{9.0cm}{\small Notes: N3198 observed in HI
by Begeman (1987), and the other galaxies in optical emission
lines by Rubin et al. (1982). \Ho{50}, and
$M/L=30$ for all galaxies. The plateau velocity, $V_p$, is
given in km/s, and $R_{25}$ and $r_o$ in kpc.}
}

\vfill

\eject
\begin{figure}

{\noindent \bf Captions to the Figures \\ }

\caption{A typical flat rotation curve can be explained by assuming
galaxy luminous matter with solar neighborhood M/L plus a dark halo.
Alternatively, it can also be explained with luminous matter having
a high M/L and test particle circular velocities given by AMOD.
}

\caption{The filled squares are the observed HI circular velocities
for NGC3198 (Begeman 1987). The solid line is the AMOD fit to the
observations with $\beta = -0.92$, $r_o = 36$ kpc, and $M/L=30$.
The galaxy mass distribution is modeled with a Plummer model (see text).
}

\caption{The squares represent the observations by Rubin et al.
(1982), and the line is the AMOD fit to the rotation curve. The
relative error in the AMOD index ($\delta A$) is given in
each frame.
}

\caption{The solid line is given by equation (6), i.e., the Tully-Fisher
relation in terms of the AMOD parameters. The dashed lines show
the 30\% error bar strip. The points were obtained from the fits to
the rotation curves shown in Figure 3, and follow the general
trend given by the predicted curve.
}

\end{figure}

\end{document}